\documentclass[11pt,a4paper]{article}
\usepackage{jcappub}

\title{\boldmath On the cosmological solutions in Weyl geometry}
\author[a,1]{V. A. Berezin\note{Corresponding author.}}
\author[a]{V. I. Dokuchaev}
\author[a]{Yu. N. Eroshenko}
\author[a]{and A.\,\,L.\,\,Smirnov}
\affiliation[a]{Institute for Nuclear Research, Russian Academy of Sciences, \\
	60th October Anniversary Prospect 7a, 117312 Moscow, Russia}
\emailAdd{berezin@inr.ac.ru}
\emailAdd{dokuchaev@inr.ac.ru}
\emailAdd{eroshenko@inr.ac.ru}
\emailAdd{smirnov@inr.ac.ru}

\abstract{We investigated the possibility of construction the homogeneous and isotropic cosmological solutions in Weyl geometry. We derived the self-consistency condition which ensures the conformal invariance of the complete set of equations of motion. There is the special gauge in choosing the conformal factor when the Weyl vector equals zero. In this gauge we found new vacuum cosmological solutions absent in General Relativity. Also, we found new solution in Weyl geometry for the radiation dominated universe with the cosmological term, corresponding to the constant curvature scalar in our special gauge. Possible relation of our results to the understanding both dark matter and dark energy is discussed.}

\begin{document}
	\maketitle
	\flushbottom

\section{Introduction and motivation}

The idea of the local conformal invariance as the fundamental symmetry of Nature come back to 1919 when Hermann Weyl made an attempt to construct the unified field theory of electromagnetic and gravitational interactions \cite{Weyl}. He noticed that Maxwell equations outside the sources are conformal invariant and claimed that the gravitational equations must have the same property. Despite the famous Einstein's criticism, Weyl conformal gravitational theory remains the remarkable generalization of Riemannian geometry if one forgets about  the electromagnetic nature of the vector field introduced by Hermann Weyl and consider it as just the gauge field mediating the conformal factors at different space-time points.

Nowadays, the same idea is advertising by Roger Penrose  \cite{Penrose,Penrose2} and Gerard 't Hooft \cite {Hooft,Hooft2} in the light of the creation of Universe from ``nothing''.

Our motivation is the following. It is well known that the requirement of the conformal invariance in the Riemanian quadratic gravity leads to the absence of the non-vacuum cosmological (homogeneous and isotropic solutions) \cite{bde16}. The gravitational theory based on the Weyl geometry, also being quadratic in curvatures, depends on several arbitrary constants (and not on their specific combinations). May be in this case cosmological solutions are allowed. And one more thing, the appearance of the new vector field incorporated into the geometry would lead to the repulsive forces between identical particles interacting with it. Interpreted as the particles of the dark matter, they will form halos around galaxies. 

In this paper we do not intend to construct the theory that fits of the observational data. We just want demonstrate what could happen if the conformal invariance appeared to be the fundamental property of Nature.

\section{Weyl geometry}

It is well known that the differential geometry of the (differentiable) manifolds is completely described by metric tensor $g_{\mu\nu}(x)$ and connections  $\Gamma^\lambda_{\mu\nu}(x)$ (not tensor). The former defines the interval, $ds$, between the neighboring points,
\begin{equation}
	ds^2=g_{\mu\nu}(x)dx^\mu dx^\nu\,, \quad (g_{\mu\nu}=g_{\nu\mu})\,,
	\label{g} 
\end{equation}
while the latter is needed for defining the parallel transfer and covariant derivatives, $\nabla_\lambda$. The parallel transfer of a vector along a closed curve around some point $x$ introduces the curvature tensor
\begin{equation}
	R^{\mu}_{\phantom{\mu}\nu\lambda\sigma}=\frac{\partial \Gamma^\mu_{\nu\sigma}}{\partial x^\lambda}-\frac{\partial \Gamma^\mu_{\nu\lambda}}{\partial x^\sigma}+\Gamma^\mu_{\varkappa\lambda}\Gamma^\varkappa_{\nu\sigma}-\Gamma^\mu_{\varkappa\sigma}\Gamma^\varkappa_{\nu\lambda}
	\label{curvaturetensor} 
\end{equation}
together with its convolutions, Ricci tensor 
$R_{\mu\nu}=R^{\lambda}_{\mu\lambda\nu}$ and curvature scalar $R=R^{\lambda}_\lambda
$. Note, that, by definition, $R^{\mu}_{\phantom{\mu}\nu\lambda\sigma}=-R^{\mu}_{\phantom{\mu}\nu\sigma\lambda}$.

It appears that the connections $\Gamma^\lambda_{\mu\nu}$ are unambiguously determined if one knows three tensors, the metric  $g_{\mu\nu}$, torsion  $S^\lambda_{\mu\nu}$  and nonmetricity $Q_{\lambda\mu\nu}$ 
\begin{equation}
	S^\lambda_{\mu\nu}=(\Gamma^\lambda_{\mu\nu}-\Gamma^\lambda_{\nu\mu})\,,
	\label{S} 
\end{equation}
\begin{equation}
	Q_{\lambda\mu\nu}=\nabla_\lambda g_{\mu\nu}.
	\label{Q} 
\end{equation}
Namely, 
\begin{equation}
	\Gamma^\lambda_{\mu\nu}=C^\lambda_{\mu\nu}+K^\lambda_{\mu\nu}
	+L^\lambda_{\mu\nu}\,,
	\label{Gamma} 
\end{equation}
where
\begin{equation}
	C^\lambda_{\mu\nu}= \frac{1}{2}g^{\lambda\kappa}(g_{\kappa\mu,\nu}+g_{\kappa\nu,\mu}-g_{\mu\kappa,\nu})
	\label{C} 
\end{equation}
is the Christoffel symbols, $g^{\lambda\kappa}$  is the inverse metric tensor ($g^{\lambda\kappa}g_{\kappa\mu}=\delta^\lambda_\mu$), and ``,'' denotes the conventional partial derivative,
\begin{equation}
	K^\lambda_{\phantom{1}\mu\nu}=-\frac{1}{2}(S^\lambda_{\phantom{1}\mu\nu}-S^{\phantom{1}\lambda}_{\mu\phantom{1}\nu}-S^{\phantom{1}\lambda}_{\nu\phantom{1}\mu})
	\label{K} 
\end{equation}
is the so called torsion tensor, and
\begin{equation}
	L^\lambda_{\mu\nu}=\frac{1}{2}(Q^\lambda_{\phantom{1}\mu\nu} -Q^{\phantom{1}\lambda}_{\mu\phantom{1}\nu}-Q^{\phantom{1}\lambda}_{\nu\phantom{1}\mu})\,.
	\label{L} 
\end{equation}

The familiar Riemannian geometry characterized by $S^\lambda_{\phantom{1}\mu\nu}=0$ and $Q_{\lambda\mu\nu}=0$. Then, the connections are just the Christoffel symbols, and the whole geometry is completely defined by the metric tensor $g_{\mu\nu}$. The curvature tensor $R_{\mu\nu\lambda\sigma} =g_{\mu\kappa}R^\kappa_{\nu\lambda\sigma}$, obeys the additional algebraic symmetries
\begin{equation}
	R_{\mu\nu\lambda\sigma}=R_{\lambda\sigma\mu\nu} = -R_{\nu\mu\lambda\sigma}=-R_{\mu\nu\sigma\lambda}\,,
	\label{symm} 
\end{equation}
\begin{equation}
	R_{\mu\nu\lambda\sigma}+R_{\mu\sigma\nu\lambda}+R_{\mu\lambda\sigma\nu} =0\,,
	\label{symm2} 
\end{equation}
and Bianchi identities
\begin{equation}
	R^\mu_{\nu\lambda\sigma;\kappa}+R^\mu_{\nu\kappa\lambda;\sigma} + R^\mu_{\nu\sigma\kappa;\lambda}=0\,.
	\label{Bianchi} 
\end{equation}
Besides, the Ricci tensor is symmetric, $R_{\mu\nu}=R_{\nu\mu}$.

Formally, Weyl geometry differs from  the Riemannian one by that now the nonmetricity tensor $Q_{\lambda\mu\nu}$ is nonzero. Namely 
\begin{equation}
	Q_{\lambda\mu\nu}\stackrel{\mathrm{def}}{=} \nabla_\lambda g_{\mu\nu}=A_\lambda g_{\mu\nu}
	\label{WeylQ} 
\end{equation}
and the connections are still symmetric, i.\,e., $S^\lambda_{\mu\nu}=0$.

Hermann Weyl (1919) considered $A_\mu$ as electromagnetic vector. We will consider it as the part of geometry under the name ``Weyl vector''.

For $\Gamma^\lambda_{\mu\nu}$ one has now
\begin{equation}
	\Gamma^\lambda_{\mu\nu}=C^\lambda_{\mu\nu}+W^\lambda_{\mu\nu}\,,
	\label{GammaW} 
\end{equation}
\begin{equation}
	W^\lambda_{\mu\nu}=-\frac{1}{2}(A_\mu \delta^\lambda_\nu + A_\nu \delta^\lambda_\mu - A^\lambda g_{\mu\nu})\,.
	\label{WW} 
\end{equation}
Note, that the curvature tensor is loosing almost all the algebraic symmetries mentioned above. Instead,
\begin{equation}
	R_{\mu\nu\lambda\sigma}=-R_{\nu\mu\lambda\sigma}-F_{\lambda\sigma}g_{\mu\nu}
	\label{RW} 
\end{equation}
and 
\begin{equation}
	R_{\mu\nu}=R_{\nu\mu}+2F_{\nu\mu}\,,
	\label{RW2} 
\end{equation}
where
\begin{equation}
	F_{\mu\nu}=A_{\nu,\mu} - A_{\mu,\nu}=\nabla_\mu A_\nu - \nabla_\nu A_\mu\,.
	\label{FW} 
\end{equation}

It is not a whole story about the Weyl geometry. Hermann Weyl noticed that the electromagnetic theory is conformal invariant outside sources (at that time particles are considered as the singularities of the space-time and are not taken into account in the theory of fields themselves). For this to be true it is sufficient to have the traceless energy-momentum tensor, and for the electromagnetic fields outside the sources it is just the case. Actually, he claimed that the conformal invariance is the fundamental symmetry in Nature.

The local conformal transformation
\begin{equation}
	ds^2=\Omega^2(x)d\hat s^2=  
	g_{\mu\nu}dx^\mu dx^\nu=\Omega^2(x)\hat g_{\mu\nu}dx^\mu dx^\nu
	\label{conf} 
\end{equation}
changes the units of the space-time measurements, but does not transform the coordinates. Under the local conformal transformation the Christoffel  symbols $C^\lambda_{\mu\nu}$ are transformed as follows
\begin{equation}
	C^\lambda_{\mu\nu}=\hat C^\lambda_{\mu\nu} +\left(\frac{\Omega_{,\mu}}{\Omega}\delta^\lambda_\nu +\frac{\Omega_{,\nu}}{\Omega}\delta^\lambda_\mu -\hat g^{\lambda\kappa}\frac{\Omega_{,\kappa}}{\Omega}\hat g_{\mu\nu}\right)\,.
	\label{confC} 
\end{equation}	
Evidently, assuming
\begin{equation}
	A_\mu=\hat A_\mu+2\frac{\Omega_{,\mu}}{\Omega}\,, 	
	\label{confA} 
\end{equation}
one readily has 
\begin{equation}
	\Gamma^\lambda_{\mu\nu}=\hat\Gamma^\lambda_{\nu\mu}\,.	
	\label{confGamma} 
\end{equation}
Therefore,
\begin{equation}
	R^\mu_{\phantom{1}\nu\lambda\sigma}=
	\hat R^\mu_{\phantom{1}\nu\lambda\sigma}, \quad R_{\mu\nu}=\hat R_{\mu\nu}, \quad F_{\mu\nu}=\hat F_{\mu\nu}\,,
	\label{confGamma} 
\end{equation}
and $A(x)$ becomes the gauge field mediating the conformal factor $\Omega(x)$ at different points.

It is in this very sense that the Weyl geometry can be considered as the conformal invariant geometry.

\section{Field equations}

Following H.Weyl, we write the simplest possible (in $4$-dim) gravitational action integral as
\begin{equation}
	S_{\rm  W}=\int\!{\cal L_{\rm W}}\sqrt{-g}\,d^4x\,,
	\label{ActionW} 
\end{equation}
\begin{equation}
	{\cal L_{\rm  W}}=\alpha_1  R_{\mu\nu\lambda\sigma}R^{\mu\nu\lambda\sigma}
	+\alpha_2R_{\mu\nu}R^{\mu\nu}+\alpha_3R^2+\alpha_4 F_{\mu\nu}F^{\mu\nu}\,.
	\label{LagrW} 
\end{equation}
Its conformal invariance is evident, since $\sqrt{-g}=\Omega^4\sqrt{-\hat g}$. The dynamical variables are $g_{\mu\nu}(x)$ and $A_\mu(x)$. The calculation of the variation $\delta S_{\rm  W}$ is extremely cumbersome, so we present here only the final result for $\delta A_\mu(x)$: 
\begin{eqnarray}
	\delta S_{\rm  W}&=&-2\!\int\biggl\{\alpha_1\left\{2(\nabla_\lambda R_{\alpha'\lambda'}) + (\nabla_\lambda F_{\lambda'\alpha'})\right\} \nonumber \\ 
	&+&\!\bigl\{(\nabla_\lambda R_{\alpha'\lambda'}) +\frac{1}{2} (\nabla_{\alpha'} R_{\lambda\lambda'}) +(\nabla_\lambda F_{\lambda'\alpha'}) \bigr\} 
	\\ 
	+&&\!\!\!\!\!\!\! 3\alpha_3(\nabla_{\alpha'} R_{\lambda\lambda'})\! + \! 2\alpha_4(\nabla_\lambda F_{\lambda'\alpha'}) \biggr\}g^{\alpha\alpha'}\!\! g^{\lambda\lambda'}(\delta A_\alpha)\!\!\sqrt{-g}\,d^4x\,. \nonumber 
\end{eqnarray}
We do not present here the results of the variation with respect to  $g_{\alpha\beta}$, because we are interested now in the cosmological solutions only, i.\,e., we are seeking for the homogeneous and isotropic space-times. In such a very special case there exists very elegant way to obtain the desired  field equation which will be described a little bit later. 

The total action integral 
\begin{equation}
	S_{\rm tot}=S_{\rm W}+S_{\rm m}\,,
	\label{Stot} 
\end{equation}
where $S_{\rm m}$ describes matter fields, its dynamical variables are some collective variable $y$, and, of course, $g_{\alpha\beta}$ and $A_\alpha$. Important note: the action integral $S_{\rm m}$, does not need to be conformal invariant, because the invariance of the gravitational action does not apply the invariance of the total action. But, the situation with the variations is completely opposed since we know exactly that the sum of the variations equals zero. One has
\begin{equation}
	\delta S_{\rm m}\stackrel{\mathrm{def}}{=}\! -\frac{1}{2}\!\int\! T^{\mu\nu}(\delta g_{\mu\nu})\sqrt{-g}\,d^4x 
-\!\!\int\!G^\mu(\delta A_\mu)\sqrt{-g}\,d^4x
+\!\!\int\!\frac{\delta \cal L_{\rm  m}}{\delta y} (\delta y)\sqrt{-g}\,d^4x=0\,.
	\label{deltaSm} 
\end{equation}
In this equation $T^{\mu\nu}$ is the energy-momentum tensor, ${\cal L_{\rm  m}}$ is the matter Lagrangian and  and $G^\mu=\delta {\cal L_{\rm  m}}/\delta A_\mu$. As usual,
\begin{equation}
	\frac{\delta \cal L_{\rm  m}}{\delta y}=0
	\label{Psi}  
\end{equation}
gives us the Euler-Lagrange equations of motion for the matter fields. Therefore, in  order $\delta S_{\rm m}$ to be conformal invariant, the following relation must be fulfilled:
\begin{equation}
	\int\!\left\{T^{\mu\nu}\left(\frac{\delta g_{\mu\nu}}{\delta\Omega}\right) +2G^\mu \left(\frac{\delta A_\mu}{\delta\Omega} \right)\right\}(\delta\Omega) \sqrt{-g}\,d^4x=0\,.
	\label{delta4} 
\end{equation}
Since
\begin{equation}
	\delta g_{\mu\nu}=2\Omega\hat g_{\mu\nu}(\delta\Omega) =2g_{\mu\nu}\frac{\delta\Omega}{\Omega}\,,
	\label{delta2} 
\end{equation}
\begin{equation}
	\delta A_\mu= 2\delta\left(\frac{\Omega_{,\mu}}{\Omega}\right) = 2(\delta \log\Omega)_{,\mu}\,,
	\label{delta3} 
\end{equation}
we obtain
\begin{equation}
	2G^\mu_{;\mu}=Trace[ T^{\mu\nu}]\,,
	\label{trace} 
\end{equation}
what can be called ``the self-consistency condition''. Here the semicolon ``;'' denotes the covariant derivative with the metric connections --- Christoffel symbols. This relation should be added to the gravitational field equations. In the Riemannian geometry, where $G^\mu\equiv0$, it is reduced to the familiar requirement  for the energy-momentum tensor to be traceless if one imposes the conformal invariance of the whole theory.

\section{Perfect fluid in Weyl geometry}

Having in mind cosmological applications we have chosen the perfect fluid as the matter content. 

Let us first consider a single particle moving in the given gravitational field. In Riemannian geometry, the only possible (differential) invariant for the particle with mass $m$ is the interval $ds$ between the neighboring points of its trajectory. Hence, the action integral equals (see, e.\,g., \cite{LL})
\begin{equation}
	S_{\rm part} = -m\!\int\!ds=-m\!\int\!\sqrt{g^{\mu\nu}(x)\frac{d x^\mu}{d\tau}\frac{d x^\nu}{d\tau}}d\tau\,.
	\label{trace} 
\end{equation}
Here the dynamical variable is the trajectory $x^\mu(\tau)$ as a function of the proper time $\tau$. It is well known that the particle moves along a geodesics --- the shortest interval.

In the Weyl geometry the situation is more subtle. There exist yet another invariant, $B$:
\begin{equation}
	B=A_\mu u^\mu,
	\label{B} 
\end{equation}
and now we are able to modify the particle action integral as follows:
\begin{equation}
S_{\rm part}=\!\int\!\!f_1(B)ds +\!\!\int\!\!f_2(B)d\tau \sqrt{-g}\,d^4x
=\! \int\!\!{\left\{\right.}f_1(B)\sqrt{g_{\mu\nu}u^\mu u^\nu} + f_2(B){\left.\right\}}d\tau\,.
	\label{Spart} 
\end{equation}
The corresponding equations of motion are
\begin{equation}
f_1(B)u_{\lambda;\mu}u^\mu =\left\{ (f_1^{'}(B)+f_2^{''}(B))A_\lambda- f_1^{'}(B)u_\lambda\right\}\!B_{,\mu}u^\mu
+\! (f_1^{'}(B)+f_2^{'}(B))F_{\lambda\mu}u^\mu\,.
	\label{B} 
\end{equation}
Since $F_{\lambda\mu}u^\lambda u^\mu\equiv0$ and $u_{\lambda;\sigma}u^\lambda\equiv0$, we obtain the condition which must be satisfied by the functions of this new invariant: either $(f_1^{'}+f_2^{''})B-f_1^{'}=0$, or $B_{,\mu}u^\mu=0$. The latter means that along the real trajectory $B=const$.

Let us turn now to to the perfect fluid. In the Riemannian geometry the action integral for the perfect fluid can be written in the form \cite{Ray,Berezin,Ber14}
\begin{eqnarray}
	S_{\rm m} &=&  -\!\!\int\!\varepsilon(X,n)\sqrt{-g}\,d^4x + \!\!\int\!\lambda_0(u_\mu u^\mu-1)\sqrt{-g}\,d^4x \nonumber \\ 
	&&+\!\!\int\!\lambda_1(n u^\mu)_{;\mu}\sqrt{-g}\,d^4x + \!\!\int\!\lambda_2 X_{,\mu}u^\mu\sqrt{-g}\,d^4x\,.
\end{eqnarray}
The dynamical variables are $n(x)$ --- the invariant particle number density, $u^\mu(x)$ --- four-velocity vector and $X(x)$ --- auxiliary variable, numbering the trajectories, $\varepsilon(X,n)$ is the invariant energy density, $\lambda_i(x)$ are Lagrange multipliers. There is no needs to write down equations of motion and all that. What we do need --- the expression for the energy-momemtum density tensor $T^{\mu\nu}$:
\begin{equation}
	T^{\mu\nu}=(\varepsilon+p)u^\mu u^\nu -p g^{\mu\nu}\,,
	\label{PerfectT} 
\end{equation}
where $p$ is the hydrodynamical pressure, 
$p=n\partial\varepsilon/\partial n -\varepsilon$.

How to incorporate our new invariant $B$ into the ``old'' perfect fluid Lagrangian? First of all, the form of the equation \eqref{Spart} suggests that our functions $f_1(B)$ and $f_2(B)$ for the single  particle are, actually, associated with particle number density $n$. Therefore,
$\varepsilon(X,n)\rightarrow  \varepsilon(X,\varphi(B) n)$. Second, there is a problem with the particle number conservation law $(n u^\mu)_{;\mu}=0$ which is one of the constraints in the conventional hydrodynamics. Indeed,  by construction,
\begin{equation}
	n=\frac{\hat n}{\Omega^4}\,,  \;\; u^\mu=\frac{\hat u^\mu}{\Omega}\,,  \;\; \sqrt{-g}=\Omega^4 \sqrt{-\hat g}\,,
	\label{nu3} 
\end{equation}
then
\begin{eqnarray}
	0&=&(n u^\mu)_{;\mu}\sqrt{-g}=(n u^\mu\sqrt{-g})_{,\mu}=\left(\frac{\hat n}{\Omega}  \hat u^\mu\sqrt{-\hat g}\right)_{,\mu} \nonumber \\ 
	&=&\frac{1}{\Omega}\left((\hat n\hat u^\mu\sqrt{-\hat g})_{,\mu}-\frac{\Omega_{,\mu}}{\Omega}\hat u^\mu\sqrt{-\hat g} \,\right).
	\label{nu4} 
\end{eqnarray}
But, $(\hat n u^\mu\sqrt{-\hat g})_{,\mu}=0$ as well, because one can just count the number of particles! Thus, we are forced to introduce some particle creation law and write down the corresponding constraint in the form
\begin{equation}
	(\varphi_1(B)n u^\mu)_{;\mu} n u^\mu)_{;\mu}-\Phi(B,n)=0
	\label{varphi1B} 
\end{equation}
(the function $\varphi_1(B)$ appeared here in order to get more freedom, just in case). Eventually, the action integral for the perfect fluid in Weyl geometry looks now as follows:
\begin{eqnarray}
	S_{\rm m}&=&-\!\!\int\!\!\varepsilon(X,\varphi(B)n)\sqrt{-g}\,d^4x + \!\!\int\!\lambda_0(u_\mu u^\mu-1)\sqrt{-g}\,d^4x \nonumber \\ 
	&+&\int\lambda_1\left((\varphi_1(B)n u^\mu)_{;\mu}-\Phi(B,n)\right)\!\sqrt{-g}\,d^4x
	\nonumber \\ 
	&+&\int\!\lambda_2 X_{,\mu}u^\mu\sqrt{-g}\,d^4x\,.
\end{eqnarray}
The set of equations of motion plus constraints are
\begin{eqnarray}
	&&-\left(1-B\left(\frac{\varphi'}{\varphi} -\frac{\varphi'_1}{\varphi_1}\right)\right) (\varepsilon+p)u_\mu +\left(\frac{\varphi'}{\varphi} -\frac{\varphi'_1}{\varphi_1}\right)(\varepsilon+p)A_\mu \nonumber \\ 
	&&+\lambda_1\frac{\partial\Phi}{\partial B} (Bu_\mu-A_\mu) -\lambda_1\left(1+B\frac{\varphi'_1}{\varphi_1}\right) n\frac{\partial\Phi}{\partial n}u_\mu
	\nonumber \\ 
	&&+\lambda_1\frac{\varphi'_1}{\varphi_1}n\frac{\partial\Phi}{\partial n}A_\mu -n\varphi'_1\lambda_{1,\mu}+\lambda_2 X_{,\mu}=0\,,
\end{eqnarray}
\begin{equation}
	-\frac{\partial\varepsilon}{\partial X}-(\lambda_1 u^\mu){;\mu}=0\,, 
	\label{varepsilon0} 
\end{equation}
\begin{equation}
	u^\mu u_\mu=1\,, \;\; (\varphi_1n u^\mu)_{;\mu}=\Phi(B,n)\,, \;\; X_{,\mu}u^\mu=0\,.
	\label{u5} 
\end{equation}
\begin{equation}
	G^\mu= \left\{\!\left(\frac{\varphi'}{\varphi} -\frac{\varphi'_1}{\varphi_1}\right) (\varepsilon+p) +\lambda_1\frac{\partial\Phi}{\partial B}
	-\lambda_1 n\frac{\varphi'_1}{\varphi_1}\frac{\partial\Phi}{\partial n}\right\}u^\mu\,,
	\label{G3} 
\end{equation}
\begin{eqnarray}
	T^{\mu\nu}&=& {\Biggl\{ }\!\left(1-B\left(\frac{\varphi'}{\varphi} -\frac{\varphi'_1}{\varphi_1}\right)\right) (\varepsilon+p) -\lambda_1B\frac{\partial\Phi}{\partial B}
	\\ 
	&\!\!\!+&\!\!\lambda_1 n\left(1\!+\!B\frac{\varphi'_1}{\varphi_1}\right) \frac{\partial\Phi}{\partial n}{\Biggl\}}u^\mu u^\nu\! +\!(-p\!+\!\lambda_1\Phi\! -\!\lambda_1 n\frac{\partial\Phi}{\partial n})g^{\mu\nu}\,.  \nonumber
\end{eqnarray}

\section{Cosmology}

By ``Cosmology'' we will understand the homogeneous and isotropic solutions, i.\,e., the Robertson-Walker metric
\begin{equation}
	ds^2=dt^2-a^2(t)\left(\frac{dr^2}{1-kr^2}+r^2(d\theta^2+\sin^2\theta d\varphi^2) \right).
	\label{RW} 
\end{equation}
Due to this, $A_\mu$ may have only one nonzero component $A_0=A(t)$, hence 
$F_{\mu\nu}=0$. Moreover, by suitable transformation with $\Omega=\Omega(t)$ one can make $A(t)=0$. We will call such solutions ``the basic solutions''. The energy-momentum tensor $T_\mu^\nu$ has the diagonal form 
$T_\mu^\nu=(T^0_0,T^1_1=T^2_2=T^3_3)$. In such a gauge all the functions of our new invariant $B$ are turned into a  set of constants.

It is easy now to drive the field equations. Surely, we are not  allowed to put $A=0$ straight into the Lagrangian before the variation, because $\delta A\neq0$. But we can do this in the resulting equations. One gets
\begin{equation}
	-6\gamma\dot R=G^0\,, \; R=-6\left(\frac{\ddot a}{a}+\frac{\dot a^2\!+\!k}{a^2}\right)\,, \; \gamma=\frac{1}{3}(\alpha_1\!+\! \alpha_2\!+\!3\alpha_3)
	\label{alpha123} 
\end{equation}
(remember, that now curvatures are Riemannian). Assuming $\Phi=\Phi_0(n)+B\Phi_1(n)+\ldots$ and denoting 
$\frac{\partial\Phi}{\partial B}= \Phi_1(n)$, we have	
$(\partial\Phi/\partial B)(0,n)= \Phi_1(n)$, we have
\begin{equation}
	G^0= -\left\{\!\left(\frac{\varphi'}{\varphi} -\frac{\varphi'_1}{\varphi_1}\right) (\varepsilon+p) 
	+\lambda_1 \frac{\varphi'_1}{\varphi_1}n\frac{\partial\Phi_0}{\partial n} +\lambda_1\Phi_1\right\}.
	\label{G0} 
\end{equation}
($\varphi,\varphi_1,\varphi',\varphi_1'$ are constants). The remaining field equations are 
\begin{subequations}\label{T00b}
	\begin{align}
		\label{eq:y:1}
	-12\gamma\left\{\frac{\dot a}{a}\dot R+ R\left(\frac{R}{12}+\frac{\dot a^2+k}{a^2}\right)\right\} &=T^0_0\,,
		\\
		\label{eq:y:2}
	-4\gamma\left\{\ddot R+2\frac{\dot a}{a}\dot R
- R\left(\frac{R}{12}+\frac{\dot a^2+k}{a^2}\right)\right\} &=T^1_1\,.
	\end{align}
\end{subequations}
The self-consistency condition looks as follows
\begin{equation}
	2\frac{(G^0a^3)^{\dot{}}}{a^3}=T^0_0+3T^1_1\,.
	\label{G0a3} 
\end{equation}
The energy-momentum tensor is 
\begin{equation}
	T^0_0=\varepsilon+\lambda_1\Phi, \quad T^1_1=-p-\lambda_1 n\frac{\partial\Phi}{\partial n} +\lambda_1\Phi\,.
	\label{T01} 
\end{equation}
Note that in our case $T^{\mu\nu}$ is conservative
\begin{equation}
	\dot T^0_0+3\frac{\dot a}{a}(T^0_0-T^1_1)=0\,.
	\label{Tconserv} 
\end{equation}
The equations of motion for matter fields are reduced to
\begin{equation}
	(\varepsilon+p)+\lambda_1n\frac{\partial\Phi_0}{\partial n} +\dot\lambda_1\varphi_1 n=0\,,
	\label{eqsmot} 
\end{equation}
\begin{equation}
	\frac{(\varphi_1na^3)^{\dot{}}}{a^3}=\Phi_0(n)\,, \quad p=n\frac{\partial\varepsilon}{\partial n} -\varepsilon\,.
	\label{eqsmot23} 
\end{equation}

With this in mind, let us consider the more simple special case when the scalar curvature is constant, $R=R_0$. Then, our set of equations becomes
\begin{equation}
	G^0=0\,, \quad 	R_0=-6\left(\frac{\ddot a}{a}+ \frac{\dot a^2+k}{a^2}\right),
	\label{special} 
\end{equation}
\begin{equation}
	T^0_0=-3T^1_1=\frac{C}{a^4}\,, \quad C=const\,,
	\label{special2} 
\end{equation}
\begin{equation}
	-12\gamma R_0\left(\frac{R_0}{12}+\frac{\dot a^2+k}{a^2}\right)=T^0_0=\varepsilon+\lambda_1\Phi_0\,,
	\label{special3} 
\end{equation}
\begin{equation}
	\varepsilon+p+\lambda_1n\frac{\partial\Phi_0}{\partial n} +\varphi_1 n\dot\lambda_1=0\,, \quad \varphi_1\frac{(na^3)^{\dot{}}}{a^3}=\Phi_0\,. 
	\label{special4} 
\end{equation}
We may not worry about the equation for $G^0$ since it can be satisfied by the choice of $\Phi_1(n)$. The immediate consequence is 
\begin{equation}
	a^2+k=-\frac{R_0}{12}a^2+\frac{Q_0}{a^2}\,, \quad Q_0=const\,,
	\label{special5} 
\end{equation}
\begin{equation}
	C=-12\gamma R_0Q_0\,.
	\label{special6} 
\end{equation}

Trivial case: $C=0$ (Weyl vacuum, $T_\mu^\nu=0$). 

Then, either $Q_0=0$ or $R_0=0$. For $Q_0=0$, the solutions are the same as in General Relativity: de~Sitter ($R_0<0$), AdS ($R_0>0$) and Minkowski space-time ($=$ Milne Universe) for $R_0=0$. 

When $Q_0\neq0$ and ($R_0=0$) we got quite new vacuum solution 
\begin{equation}
	\pm(t-t_0)= \frac{1}{2}\!\int\!\frac{da^2}{\sqrt{Q_0-ka^2}}\,.
	\label{new} 
\end{equation}

Nontrivial case: $C\neq0$.

The solutions for the scale factor $a(t)$ are
\begin{equation}
	a^2=-6\frac{k}{R_0}+\!\sqrt{-12\frac{Q_0}{R_0} -36\frac{k^2}{R_0^2}}\sinh\left[\!\sqrt{-\frac{R_0}{3}}(t-t_0)\right],
	\label{nontriv1} 
\end{equation}
if  $R_0<0$,  $Q_0>-3k^2/R_0$;
\begin{equation}
	a^2=-6\frac{k}{R_0}+\!\sqrt{12\frac{Q_0}{R_0} + 36\frac{k^2}{R_0^2}}\cosh\left[\!\sqrt{-\frac{R_0}{3}}(t-t_0)\right],
	\label{nontriv2} 
\end{equation}
if $R_0<0$, $Q_0<-3k^2/R_0$;
\begin{equation}
	a^2=-6\frac{k}{R_0}\exp\left[\pm\sqrt{-\frac{R_0}{3}}(t-t_0)\right],
	\label{nontriv3} 
\end{equation}
if $R_0<0$, $Q_0=-3k^2/R_0$;
\begin{equation}
	a^2=-6\frac{k}{R_0}+\!\sqrt{-12\frac{Q_0}{R_0} + 36\frac{k^2}{R_0^2}}\sin\left[\!\sqrt{\frac{R_0}{3}}(t-t_0)\right],
	\label{nontriv4} 
\end{equation}
if $R_0>0$, $Q_0<3k^2/R_0$. 

Before coming to the end, we would like to make one important remark. When interpreting the observational cosmological data we are using Friedmann equations for the scale factor. Therefore, it would be instructive to see, what kind of the effective energy-momentum tensor $T(eff)^{\mu\nu}$ corresponds to our nontrivial solution. Let us remind Friedmann equations:
\begin{subequations}\label{friedmann}
	\begin{align}
		\label{eq:y:3}
3\frac{\dot a^2+k}{a^2}&=8\pi G T(eff)^0_0\,,
		\\
		\label{eq:y:4}
2\frac{\ddot a}{a}+\frac{\dot a^2+k}{a^2}&=8\pi G T(eff)^1_1\,.
	\end{align}
\end{subequations}
For our nontrivial solution one gets (remember $R=R_0=const$):
\begin{subequations}\label{nonfriedmann}
	\begin{align}
		\label{eq:y:5}
		8\pi G T(eff)^0_0&=3\frac{Q_0}{a^4}- \frac{R_0}{4}\,,
		\\
		\label{eq:y:6}
	8\pi G T(eff)^1_1&=-\frac{Q_0}{a^4}- \frac{R_0}{4}\,. 
	\end{align}
\end{subequations}
When $Q_0/a^4 \gg |R_0|/4$, we  obtain the effective equation of state of the radiation dominated universe. When the equality is reversed, we obtain the dark energy ($R_0<0$, corresponding to the positive cosmological constant). Two notes are in order. First, the expressions for $T(eff)^\nu_\mu$ remain valid in general gauge with $a\rightarrow a\Omega$. Second, they make sense only if some nontrivial solution does exist.

At the end we demonstrate one of the possible nontrivial solutions with constant curvature. The very existence of the presented solutions comes from the fact, tat we have, actually, 4 equations for 4 unknown functions: $\varepsilon(n)$, $n(t)$, $\Phi_0(n)$ and $\lambda_1(t)$.

Let $\varepsilon=3p$, then 
\begin{equation}
	\varepsilon =\varepsilon_0n^{4/3}\,,  \quad  \Phi_0=\varphi_0 n^{4/3}\,,
	\label{nontriv4} 
\end{equation}
\begin{equation}
	\dot\lambda_1=-\frac{4}{3}\frac{C}{\varphi_1} \frac{1}{a^4}\,,  \quad  \left((na^3)^{-1/3}\right)^{\dot{}}=-\frac{1}{3}\frac{\varphi_0}{\varphi_1}\frac{1}{a}\,.
	\label{nontriv5} 
\end{equation}

\section{Conclusions and discussions}

In this paper we investigated the possibility of construction the homogeneous and isotropic (cosmological) solutions in the Weyl gravity. 

We have chosen the quadratic Lagrangian for the conformal Weyl gravity and the perfect fluid as the matter field. It was noticed that the matter action integral is not obliged to be conformal invariant, but its variation --- does. This imposes the additional, self-consistency condition, on the matter energy-momentum tensor. It is this condition that requires the modification of the perfect fluid Lagrangian in order to include the particle creation.

We considered the action integral for single particle moving in the given gravitational field in Weyl geometry and found that there exists yet another invariant, absent in the Riemanian geometry (and, particularly, in General Relativity). Namely, it is the convolution of the Weyl vector and particle four-velocity. We also suggested some modification of the perfect fluid Lagrangian by inclusion such an invariant. 

What concerns the cosmological setup, i.\,e., homogeneous and isotropic manifold, the only unknown function in their metric  is the scale factor, depending on time coordinate, and the Weyl vector has only temporal component depending on time. In such a case it is possible to find the appropriate conformal transformation in order to put it zero  as well. The corresponding solutions  we call ``the basic solutions''. We found all the vacuum basic solutions with constant curvature, one of them is quite new (no analog in General Relativity). Also we found a non-vacuum solution when the curvature scalar is constant. The general case deserves further investigation but the very existence of the novel solutions is quite obvious. It is important to note that these solutions may have nonzero trace of the energy-momentum tensor (unlike in the case of the conformal invariance in Riemannian geometry).

The particles interacting with Weyl vector (due to our modification) may be interpreted as a dark matter. 

Let us imagine that there are two types of particles: Weyl-neutral and Weyl-charged. The former is just the ordinary, visible, matter, while the latter represents the dark one. On large, cosmological, both of them do not interact with the Weyl vector (in our special gauge). But on the galactic scale where the homogeneity and isotropy conditions are broken, the identical dark particles will experience repulsion what could explain the observed cores in the centers of some dark matter halos (e.\,g., in some dwarf galaxies).

The special gauge (with zero Weyl vector) appears to be particularly convenient for comparing with the observational data. The matter is that we are using Friedmann (general relativistic) equations when interpreting the observed phenomena, thus postulating the absence of the Weyl vector. We demonstrated this by showing explicitly how the cosmological term ($=$ dark energy) emerges due to such difference in the field equations.

\acknowledgments

We are grateful to E.\,O.\,Babichev for stimulating discussions.

\end{document}